# New method for characterization of magnetic nanoparticles by scanning magnetic microscopy

**Jefferson F. D. F. Araujo** [1,*], **Tahir** [1], **Soudabeh Arsalani** [2], **Fernando L. Freire Jr.** [1], **Gino Mariotto** [3], **Marco Cremona** [1], **Leonardo A. F. Mendoza** [4], **Cleanio Luz-Lima** [5], **Tommaso Del Rosso** [1], **Oswaldo Baffa** [2] and **Antonio C. Bruno** [1]

[1] Department of Physics, Pontifícia Universidade Católica do Rio de Janeiro, 22451-900, Rio de Janeiro, Brazil

[2] Department of Physics, FFCLRP, University of São Paulo, 14040-901, Ribeirão Preto, SP, Brazil

[3] Department of Informatics, Università di Verona, Strada le Grazie 15, I-37134 Verona, Italy

[4] Department of Electrical Engineering, Universidade Estadual do Rio de Janeiro, Rio de Janeiro 20550-900, Brazil

[5] Department of Physics, Campus Ministro Petrônio Portella, Universidade Federal do Piauí, Teresina 64.049-550, PI, Brazil

*\* Corresponding author: jferraz@fis.puc-rio.br*

**Abstract** - In this paper, we present a new method for the magnetic characterization of bulk materials, microstructures, and nanostructures. We investigated the magnetic and morphological properties of two colloidal dispersions of iron oxide ($Fe_3O_4$) magnetic nanoparticles (MNPs), synthesized by chemical precipitation (co-precipitation) and pulsed laser ablation (PLA) in liquid, by scanning magnetic microscopy (SMM) applied to a small sample with mass on the order of tens of µg. We evaluated the performance of this technique by comparing magnetization curves and measurements obtained with commercial magnetometers, considered standard. The errors obtained for the saturation and remanent magnetization were approximately ± 0.18 $Am^2/kg$ and ± 0.6 $Am^2/kg$, respectively. The average size distribution of the NPs estimated from the magnetization curve measurements is consistent with the results obtained by traditional transmission electron microscopy (TEM). The technique can be extended to measure and analyze magnetization curves (hysteresis loops), thus enabling an even more accurate estimation of overall NP sizes.

## 1. Introduction

Studies on the magnetic properties of materials have received attention in physics and engineering [1-5]. Certain important parameters of magnetic materials, such as the Curie temperature, saturation magnetization, remanent magnetization, coercive fields and magnetic anisotropy, direct their applicability. Nanoparticles (NPs) have demonstrated great promise for medical applications including magnetic resonance imaging (MRI), drug delivery, magnetic particle imaging (MPI) and hyperthermia [6-7]. The most frequently used MNPs for biomedical applications are iron oxide ($Fe_3O_4$) NPs because of the low toxicity, high magnetism and biocompatibility of these NPs compared with other NPs such as $CoFe_2O_4$, $ZnFe_2O_4$ and $MnFe_2O_4$. The physical properties of NPs including size, composition, shape and surface chemistry vary widely and influence the biological properties and clinical applications. There are many physical and chemical methods that are used for the synthesis of $Fe_3O_4$ NPs, but the most commonly used methods are PLA in liquid and co-precipitation [8-11]. In our present configuration for the production of NPs by PLA, the amount of magnetic material is on the order of tens of µg, this amount can be used, for example, in the study *In vitro* [12]; therefore, magnetic characterization is not feasible using certain magnetometry techniques [13-15]. Accordingly, we develop a technique to acquire magnetization curves from magnetic maps obtained by scanning magnetic microscopy (SMM) to characterize magnetic materials [16]. We compare the magnetization results with measurements obtained from independent commercial magnetometers, yielding errors of approximately ± 0.18 Am$^2$/kg (0.37%) at saturation and below 0.19 Am$^2$/kg (0.38%) for the remanent magnetization. From the magnetic measurements, we can determine the size of MNPs with results comparable to traditional techniques such as transmission electron microscopy (TEM).

Using a small amount of mass, this configuration enables the application of magnetic fields up to 550 mT on a sample. This approach has a scanning range of 150 to 150 mm with µm resolution. In the current configuration, the microscope is equipped with a pair of commercial Hall effect sensors that form an axial gradiometer. The output noise measured at 6.0 Hz is approximately 250 nT$_{rms}$/√Hz in an unshielded environment, and the sensitivity of the magnetic moment is 4.20 x 10$^{-11}$ Am$^2$.

## 2. Magnetic microscope

The magnetic microscope can characterize magnetic samples (bulk, liquid, micro and nanostructured) with mass on the order of µg. We placed the sample, upside down on the sample holder with double-sided adhesive tape, between the poles of the electromagnet, which is capable of generating direct current (DC) magnetic fields of up to 550 mT in a 40-mm-diameter area with a 17-mm pole gap at a current of 4.0 A. We positioned the electromagnet such that its polar axes were oriented in the vertical direction (Fig. 1(a)). To detect the response of the sample to the magnetic applied field, we used two Hall sensors, hereinafter denoted Sensor A and Sensor B, that incorporate an GaAs element in a surface-mount technology (SMT) package [16]. The detection areas were 200 µm in diameter, and we placed the sensors at a nominal distance of 130 µm from the upper surface of the sample. To demonstrate the capability of the equipment, we present the result of mapping a print positioned over a sample holder (see Fig.1(b)). In figure 1(c) shows the experimental map made in the scanning magnetic microscope, applying a field of 20 mT.

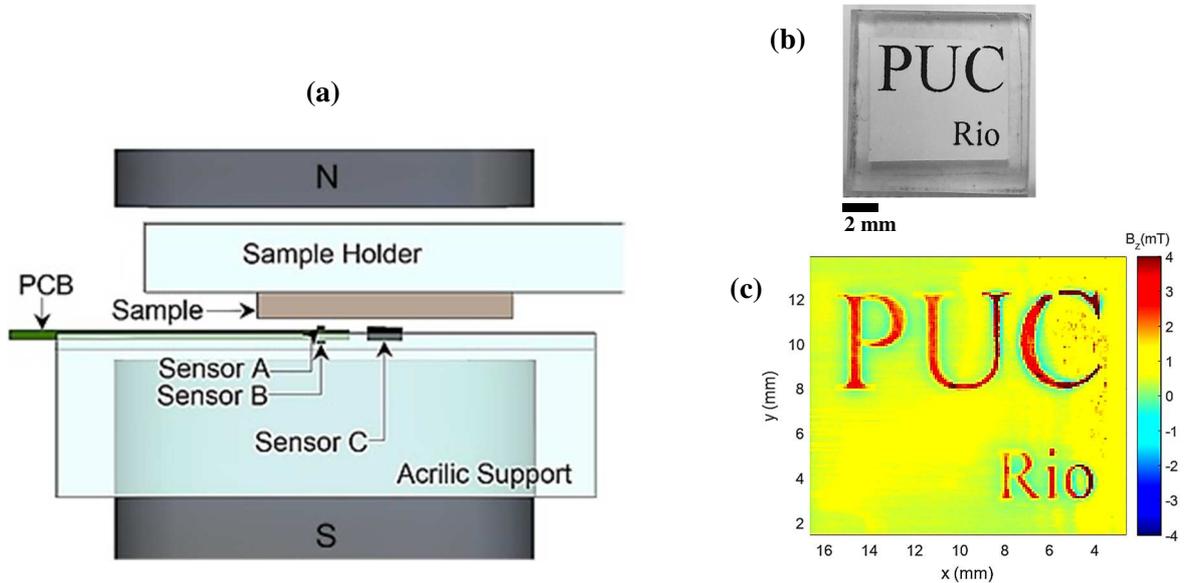

**Figure 1. (a)** Scanning magnetic microscope. **(b)** Photo of the sample holder, where it was attached the print of the name PUC-Rio. **(c)** Experimental map made in scanning magnetic microscope, applying a field of 20 mT.

Gradiometer is designed to attenuate the applied field, thereby increasing the dynamic range of the instrument and enabling operation under strong applied fields. However, the gradiometer can also attenuate ambient magnetic noise. We adjusted the field attenuation by applying a uniform field of 550 mT, and we varied the polarization current of Sensor B until the output reached 0.5 mT, corresponding to a field attenuation factor, or common mode rejection, of 1,000 [16-17].

## 3. Calibration process and sample holder selection

3.1 Microscope calibration

To obtain the voltage output of the SMM in terms of the magnetic moment of the sample, a calibration procedure must be performed to relate the output to the magnetic field detected. For the calibration source, we used magnetic microparticles ($Fe_3O_4$). We inserted 18.2 mg of these microparticles into a 3.0-mm-deep cylindrical cavity with a diameter of 3.0 mm. Due to the relationship between the $z$-component of the magnetic field $B_z(x, y, z)$ and the magnetic moment ($m_z$) when the sample is close to saturation because of the proximity of the sample to the sensor, the magnetic dipole model cannot determine the magnetic moment of the sample. Therefore, we chose a cylindrical current sheet model based on the cylindrical design of the sample holder (see Eq. 1) [17]:

$$m_z = \frac{4\pi^2 a^2 L}{\mu_0 \int_{-L/2}^{L/2} \int_0^{2\pi} \frac{(x-x_0) a \cos(\phi) d\phi}{r^3} dx} B_z(x, y, z) \quad (1)$$

where $r = \sqrt{(x - x_0)^2 + (y - a\sin(\phi))^2 + (z - a\cos(\phi))^2}$, $\mu_0$ is the permeability of free space, and $a$ is the radius of a uniformly magnetized cylinder along its length $L$. We determined the actual distance $z$ between the sample and the sensor by performing a linear scan at the top of sample $B_z = (x, y, z)$ and analyzing only the spatial dependence of $B_z$, where a = 1.5 mm and L = 3.0 mm. Using a least-squares fitting routine, we obtained a distance of 138 μm between the Hall sensor element and the top of the sample. Using this distance and a density value of 5.19 x $10^{+3}$ kg/m$^3$ for bulk $Fe_3O_4$, we obtained a magnetization of 75.8 Am$^2$/kg for 500 mT, which is approximately 0.16% greater than the value obtained from a vibrating-sample magnetometer (VSM) and 0.21% lower than the value obtained from a Hall magnetometer [18-19].

3.2 Choice of sample holder

We measured several materials to identify a sample holder with a uniform response, behavior close to paramagnetism and lower intensity in the presence of an applied magnetic field. These characteristics were important because the nanoparticle samples had small masses of approximately 60 μg and therefore exhibited a weak magnetic response in the presence of an applied field.

Figs. 2(a) - (c) show three samples composed of glass (microscope slide), quartz, and acrylic, respectively. We obtained the maps in the figure by applying a field of 500 mT in the $z$-direction (perpendicular to the sample face). Fig. 2(a) shows the $z$-component of the magnetic field generated by the glass sample after an acid bath with an induced field of approximately 3.0 mT; the field was relatively uniform throughout the sample region and had a magnitude of less than 1% of the applied field. Fig. 2(b) shows a magnetic map of the quartz sample after an alcohol bath. The induced magnetic field of this sample (Fig. 2(b)) was less uniform than that of the glass sample, although this sample had a lower intensity of 1.5 mT, which is equivalent to 0.3% of the applied field. Fig. 2(c) shows a scan of the acrylic sample after an alcohol bath, displaying a clear pattern

with low-intensity induced fields of approximately 1.0 mT, corresponding to 0.2% of the applied field, that were uniform throughout the material.

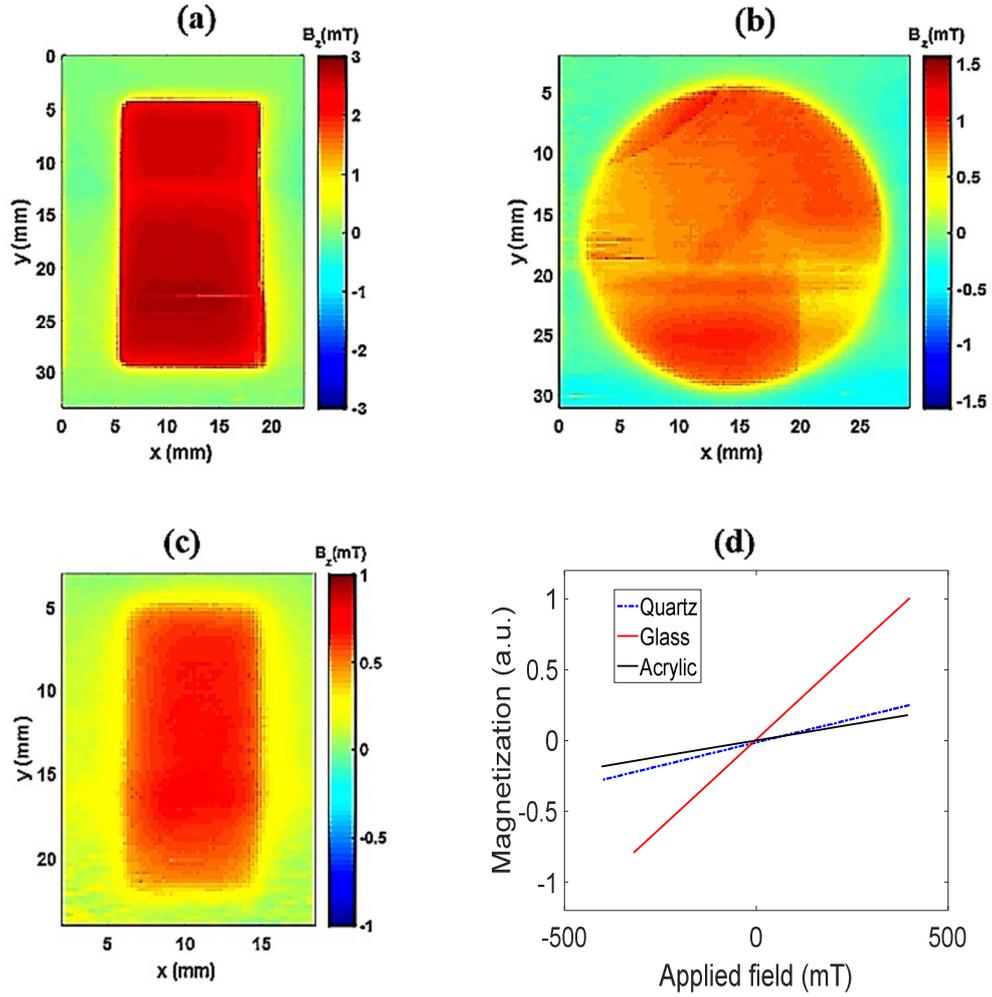

**Figure 2.** Magnetic maps of the samples at 500 mT. (a) Magnetic map of the induced field for a rectangular glass sample, 2.5 mm x 1.3 mm in size. (b) Map of a 2.5-mm-diameter quartz disc. (c) Magnetic map of the induced field for a rectangular acrylic sample, 1.7 x 0.9 mm in size. (d) The magnetization results obtained after the modeling process.

To quantify these results in terms of magnetization and to aid in the selection of a sample holder, we used theoretical models to obtain the magnetization curve of each sample holder (magnetic moment divided by the material mass). The best choice was a uniformly magnetized rectangular prism model due to the similarity of this model to the sample geometries shown in Figs. 2(a) and (c) (glass and acrylic, respectively).

The $z$-component of the magnetic field generated at any point in the $B_z$ space ($x, y, z$) for a rectangular prism with a uniform volume of magnetization $m_z$ can be written as [16]

$$\boldsymbol{B}_z(x,y,z) = \frac{\mu_0 m_z}{4\pi}[F(-x,y,z) + F(-x,y,-z) + F(-x,-y,z)$$
$$+ F(-x,-y,-z) + F(x,y,z) + F(x,y,-z) \quad (2)$$
$$+ F(x,-y,z) + F(x,-y,-z)]$$

where the F function is defined as

$$F(x,y,z) = arctan \frac{(x+a)(y+b)}{(z+c)\sqrt{(x+a)^2 + (x+b)^2 + (x+c)^2}}$$

and *a*, *b* and *c* are the half-width, half-length and half-thickness of the rectangular prism, respectively.

For the quartz sample holder shown in Fig. 2(b), we chose a cylindrical current sheet model due to the cylindrical design of the sample holder. This model was the same as the model used to calibrate the microscope (see Eq. 1). Fig. 2(d) shows the magnetization results obtained after the modeling process. All the samples exhibited paramagnetic behavior up to 500 mT; the results for the quartz and acrylic samples had the lowest slope, and thus, these samples had the lowest magnetic response in the presence of an applied magnetic field. The acrylic sample was more suitable because this sample had the weakest and constant induced magnetic field; therefore, we chose the acrylic sample holder to characterize the magnetic samples.

## 4. Experimental methods

4.1 Synthesis of the Colloidal Dispersions of Iron Oxide NPs

We obtained the iron oxide NPs by both PLA in water and co-precipitation of ferric and ferrous salts. For the synthesis of the MNPs by PLA in water, first, we removed the surface oxides from a disk of pure iron (Kurt & Lesker, U.SA, 99.995% purity) using sandpaper, and then, we placed the disk in two subsequent ultrasound baths of milli-Q water for 10 minutes. Subsequently, we rinsed the iron disk, immersed the disk in 2 mL ultrapure deionized water and irradiated the disk by ns laser pulses at a wavelength of 1064 nm with a repetition rate of 10 pulses per second using the same experimental setup described in [20]. We adjusted the lens and target distance of the optical setup to obtain laser pulses with a low fluence on the order of 0.2-0.4 J/cm$^2$. We chose this low fluence to have better control of the concentration of the colloidal dispersions of iron oxide NPs. We terminated the PLA process after approximately two hours, and then, we precipitated the NPs using magnetic separation and washed the NPs several times with deionized water. We measured the NP concentration by inductively coupled plasma mass spectrometry (ICP-MS) using a Nexion300X Perkin Elmer, USA. The total mass/concentration of iron oxide NPs synthesized in each ablation cycle was less than 65 µg /32 ± 2 µg / mL.

The co-precipitation procedure was described in [9], briefly, we first dissolved 2.25 g $FeCl_3 \cdot 6H_2O$ in 8.5 mL distilled water. Then, we dissolved 1.32 g $FeCl_2 \cdot 4H_2O$ in 3.5 mL aqueous hydrochloric acid (5.45 M). We added the mixture of iron(III) chloride (4 mL) and iron(II) chloride (1 mL) to a basic $NH_4OH$ solution (1.28 M) with vigorous stirring in a water bath at 90 °C. The color of the solution changed immediately to black, indicating the formation of iron oxide ($Fe_3O_4$) NPs. We precipitated the NPs using magnetic separation and washed the NPs several times with Milli-Q water until the solution reached a neutral pH. Finally, we dried the $Fe_3O_4$ at room temperature [9].

4.2  Raman Spectroscopy

To perform Raman analysis, we deposited the nanomaterials synthesized by both chemical and physical routes over a gold thin film to quench the luminescence of the nanomaterials following the same experimental procedures described in [20]. We carried out careful vibrational spectroscopy analysis of the precipitated nanoparticles in the form of aggregates with a granular structure using a Horiba Jobin-Yvon (model LabRam HR800) microRaman spectrometer. This system was equipped with a He-Ne laser as an excitation source ($\lambda_{exc}$ = 632.8 nm) and a notch filter for the rejection of the Rayleigh line. We focused the laser beam onto the sample surface through a 100x objective lens with a numerical aperture of 0.9, giving a spot size of approximately 1 µm. The scattered radiation was dispersed by diffraction grating of 600 lines/mm and detected at the spectrograph output by a multichannel detector, a CCD with 1024 x 256 pixels cooled by liquid nitrogen with maximum efficiency occurring in the red region. The average spectral resolution was approximately 1 cm$^{-1}$/pixel over the spectral range of interest, while the low wavenumber limit was approximately 200 cm$^{-1}$ due to the notch filter. We collected Raman spectra in the backscattering geometry over the Stokes-shifted region, and we kept the irradiation power at the surface of the sample below 100 µW to avoid any sample damage. Under these excitation conditions, very long integration times (typically 1800 s) were necessary to collect spectra characterized by an optimal signal-to-noise ratio. Micro Raman measurements were carried out from different aggregates of both kinds of NPs under the experimental conditions described above, and the spectra recorded showed very good reproducibility. Moreover, repeated measurements over the same sample grain gave overlapping spectra, thus indicating that no structural damage was caused by the laser beam focused on the sample surface.

4.3    Dynamic Light Scattering and Transmission Electron Microscopy

We initially determined the size distributions of the synthesized nanomaterials by both dynamic light scattering (DLS) and TEM techniques. We performed the DLS using a Nano SZ-100 HORIBA Scientific nanoparticle analyzer, Japan, with an incident beam at a wavelength of 532 nm. For the TEM measurements, we used a Tecnai G2 Spirit TWIN FEI microscope, USA, operating at 120 kV with a LaB$_6$ (lanthanum hexaboride) filament. To prepare the sample for TEM analysis, we deposited a drop of 20 µL colloidal solution of the nanomaterial diluted at a concentration of 8 µg / mL over a copper grid covered with a conductive polymer and allowed the grid to dry for one night in air.

**5. Results and discussion**

5.1 Raman spectroscopy

Fig. 3 shows two typical Raman spectra of the colloidal aggregates of iron oxide NPs synthesized by co-precipitation (black color) and PLA (red color) methods. In general, the observed micro Raman spectra were imposed on a continuous, nearly flat luminescence background of relatively lower intensity with respect to the main Raman bands of iron oxide. The spectra of the two precipitated NPs clearly show certain common features that seem to be related to the same iron oxide phase present in both types of NPs. However, additional bands in the spectrum of NPs obtained by the co-precipitation method suggest a more complex structural arrangement in this system, which is also reflected by the broader bandwidths of the peaks.

The spectrum of NPs synthesized by PLA is relatively simpler since it consists of three main spectral features centered at approximately 317, 548 and 670 cm$^{-1}$, which are typical signatures of magnetite (Fe$_3$O$_4$) [21-25]. Magnetite has a spinel structure belonging to space group O$_h^7$, and according to group theory, gives rise to five Raman modes: three $T_{2g}$, one $E_g$ and one $A_{1g}$ [22]. Three of these modes are clearly present in the spectrum of NPs obtained by PLA: the strongest peak at approximately 670 cm$^{-1}$, identified as the $A_{1g}$ mode, corresponds to the stretching vibrations of the oxygen atoms along the Fe-O bonds, while the weaker bands at 548 and 317 are associated with $T_{1g}$ and $E_{2g}$ modes, respectively [26]. Finally, the very broad bands above 1000 cm$^{-1}$ of relatively lower intensity are probably due to second order scattering processes.

The spectrum of MPs synthesized by the co-precipitation method shows a strong scattering component at approximately 715 cm$^{-1}$ turning out as a shoulder of the main peak at approximately 670 cm$^{-1}$. This spectral component, together with other broad features resulting from band overlap observed in the region between 400 and 600 cm$^{-1}$, is characteristic of the iron oxide maghemite [γ-Fe$_2$O$_3$], which has an inverse spinel structure and can be considered as an iron deficient form of magnetite.

In conclusion, the chemically precipitated powders have a definitively more complex structure (maghemite mixed with magnetite) than that of the NPs synthesized by PLA, where only iron oxide magnetite phase is detected by Raman spectroscopy.

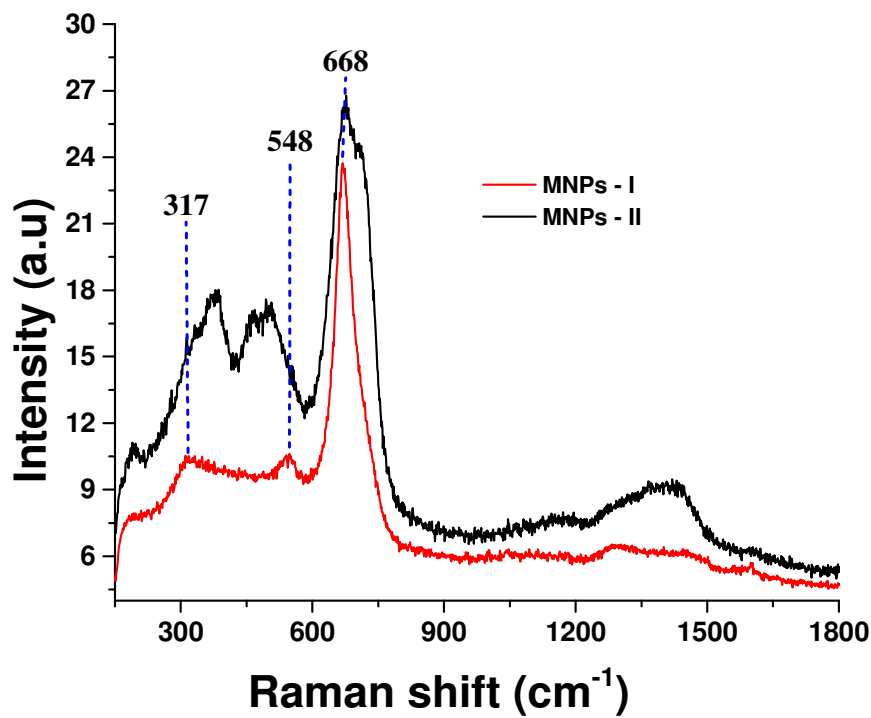

**Figure 3**. Raman spectra of the iron oxide NPs synthesized by PLA (red line) and co-precipitation (black line).

5.2 Magnetization curve

Before performing magnetic characterization, the induced magnetic field signal of the sample only must be obtained to avoid errors in the magnetization curve. In this case, a 17 mm x 17 mm structure (acrylic sample holder) was scanned. At the center there was a cylindrical cavity with a diameter of 400 µm and a depth of 400 µm; filled with ~50 µg magnetite NPs produced by co-precipitation (Fig. 4(a)). Fig. 4(b) shows a magnetic map of the perpendicular component of the magnetic field response obtained from a 2 mm x 2 mm scan of the sample holder under a field of 420 mT applied perpendicular to the sample surface. In this way, we observed the magnetic field response of the NPs and sample holder. From the magnetic map (Fig. 4(b)), we obtained the position (in XY space) of the maximum intensity and the induced field of the sample in the presence of the applied magnetic field (curve S1 in Fig. 4(c)); The S1 curve represents the induced field of the sample combined with the contribution of the sample holder, and we observed a signal in the S1 graph due to the sample holder starting at approximately 6.0 mm.

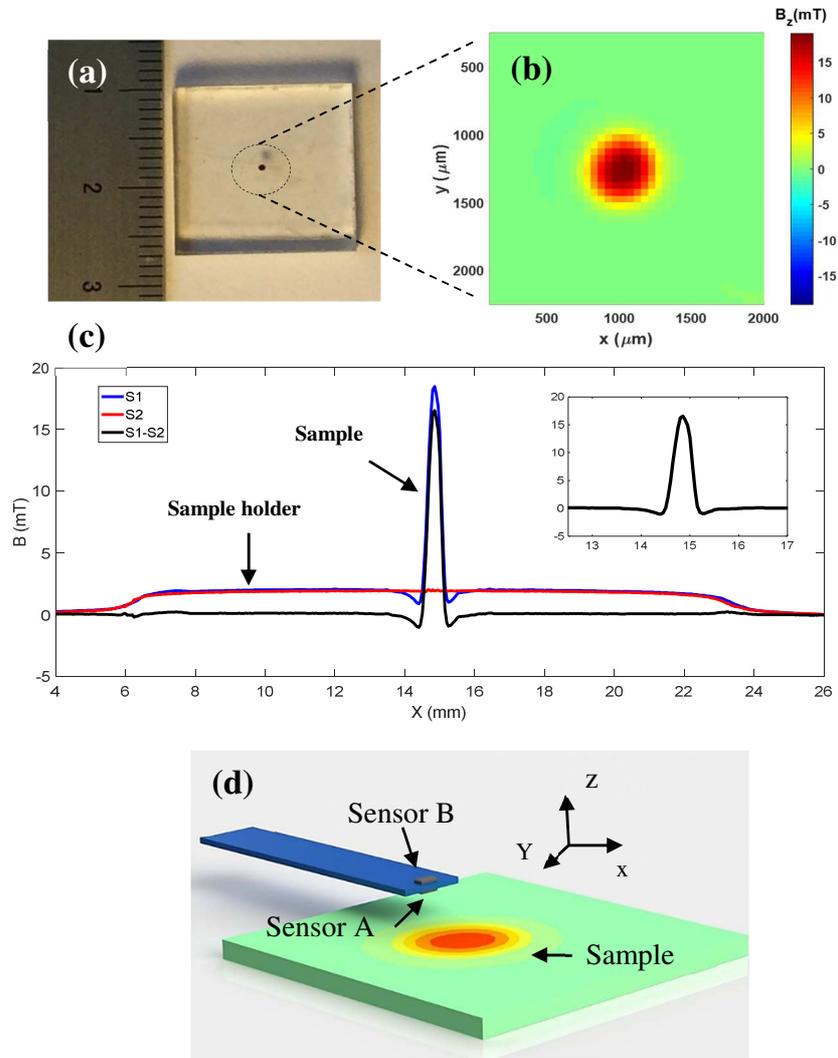

**Figure 4.** (a) Image of the acrylic sample holder, which is square-shaped with a side length of 1.7 mm, used to characterize the magnetic NPs in the center of the 400 x 400 µm cylindrical cavity. (b) Magnetic map of ~50 µg magnetic $Fe_3O_4$ NPs produced by co-precipitation and placed in the cylindrical cavity. (c) Results obtained along one axis in the presence of a 420 mT applied field. The S1 path passes through the center of the cavity, which contains 50 µg NPs, and the S2 path passes through the center of the empty cavity; S1 - S2 is the subtraction of one path from the other (this increment is the result of the S1 - S2 subtraction). (d) Spatial position of the magnetic reading system and the magnetic map shown in (a).

In this case, it was necessary to eliminate this signal. In Fig. 4(c), the S2 curve presents the results for only the sample holder without the NPs; thus, the induced field curve was obtained by subtracting curve S1 from curve S2 (Fig. 4(c) — increment, S1-S2 curve). In this way, we determined the induced field due to only the NPs. Fig. 5 shows a set of fifteen experimental maps of $B_z$ (50-µm steps) for $Fe_3O_4$ NPs.

After this process, we obtained the induced magnetic field values of the sample in the presence of each field applied by the electromagnet. With these values and using the model of a uniformly magnetized cylinder (Eq. 1), we obtained the magnetic moment of the sample. From the maps, the curves with maximum intensity are obtained. Fig. 6(a) presents such curves of the induced magnetic field in the sample; the blue curves were obtained in the presence of a positive magnetic field, and the red curves were obtained for a negative magnetic field. Fig. 6(b) demonstrates the agreement between the theoretical model (solid black curve) and the values obtained for the induced field of the sample (blue circles). Fig. 6(c) compares the magnetization results obtained using our technique (blue circles) for a 1.7 mg sample of the same NPs produced by co-precipitation with the values obtained using a commercial magnetometer (MPMS SQUID, Quantum Design Inc.) (solid red curve).

For absolute magnetization errors, see the inset in Fig. 6(c). Near saturation, the error is ± 0.18 $Am^2$/kg, and for field values close to zero, the error is approximately ± 0.6 $Am^2$/kg. In Fig. 6(d), the magnetization curves of the MNPs produced by laser ablation and co-precipitation are displayed. Both curves show superparamagnetic behavior, the curves do not exhibit remanence or coercivity.

Fig. 6(e) shows the zero-field-cooled (ZFC) and field-cooled (FC) curves of the NPs produced by PLA (MNPs — I) and co-precipitation (MNPs — II). In ZFC curves, magnetization increases with temperature. This peak temperature of a ZFC curve is identified as the blocking temperature ($T_B$), and the figure 6. (e) shows that shows that the magnetization curves at ambient temperature are above the blocking temperature.

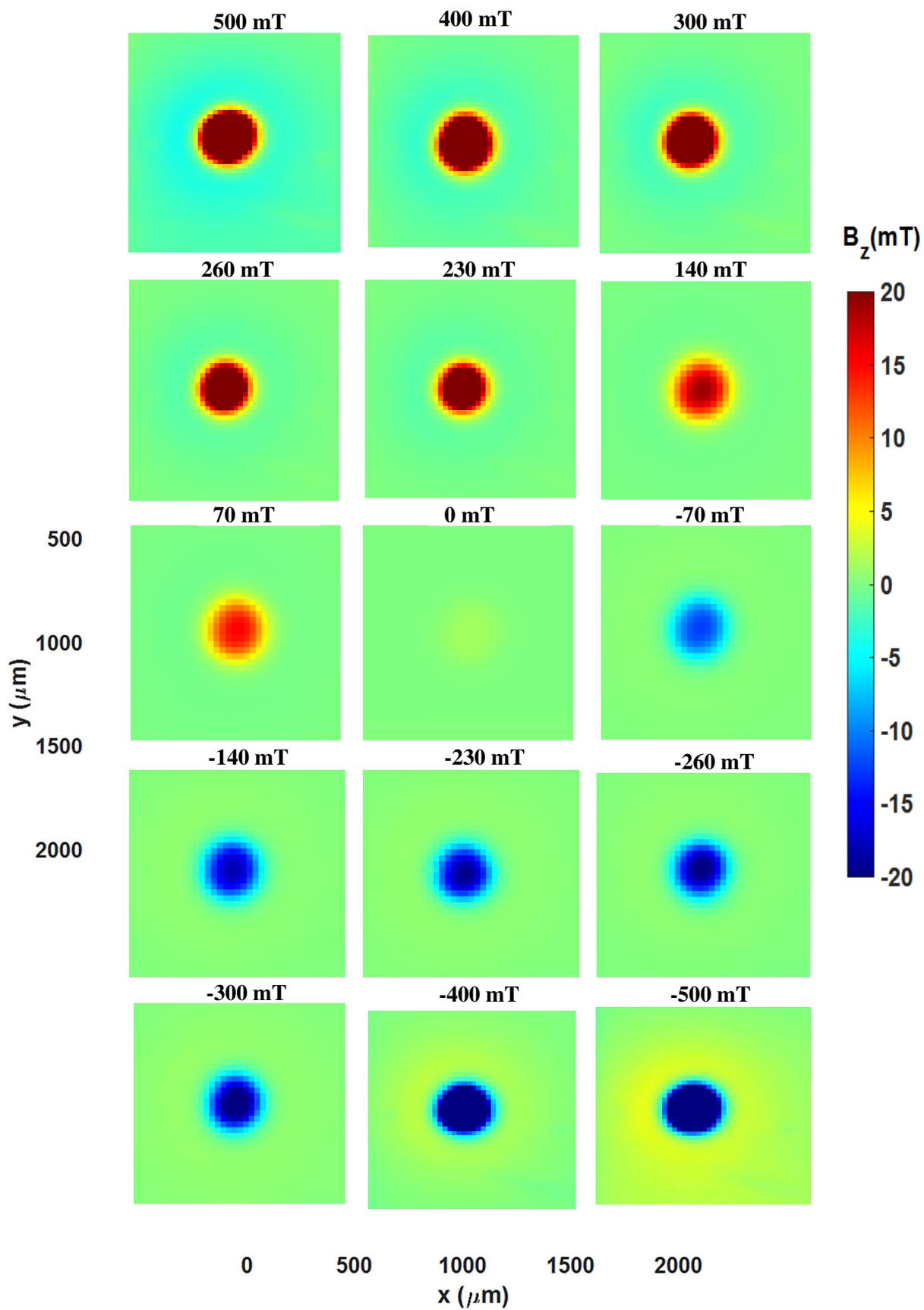

**Figure 5.** Maps of a 2.1 mm x 2.1 mm area surrounding a cylindrical cavity containing a few tens of μg of magnetic NPs as the applied field varies from 500 mT to - 500 mT.

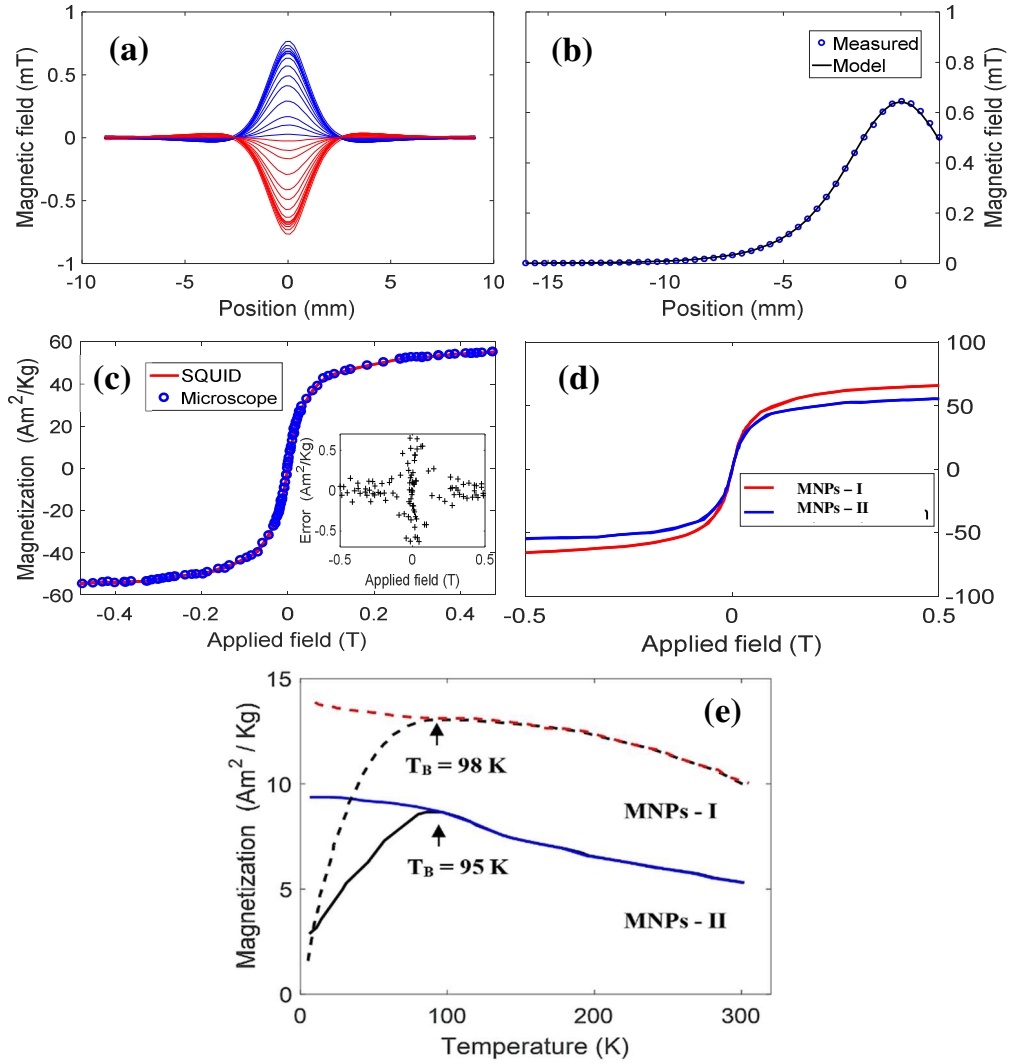

**Figure 6.** (a) Several curves of the induced magnetic field of the sample; the blue curves were obtained in the presence of a positive magnetic field, and the red curves were obtained in the presence of a negative magnetic field. (b) Comparison of the theoretical results (solid black curve) and the values obtained from the induced field of the sample (blue circles). (c) Comparison of the magnetization results obtained using our technique (blue circles) and the values obtained for the same nanoparticle sample using a commercial magnetometer (MPMS SQUID, Quantum Design Inc.) (solid red curve). (d) Magnetization curves of the samples produced by PLA (MNPs — I) and co-precipitation (MNPs — II). (e) Low temperature measurements of the MNPs.

We can also obtain the magnetic moment directly from the experimental maps (Fig. 7(a)). Using this modeling technique, we construct a theoretical model, using the model of a current cylinder (See Eq. 1), which takes into account the entire mapping region (See Fig. 7(b)). Then to verify if the modeling works we subtract the map obtained experimentally by the theoretical model, the result can be seen in figure 7(c). We can see in figure 7(c) that there is information that was not used to obtain the magnetic moment, making this method less precise than the previous modeling method (Fig. 6 (a) - (c)). An example can be seen in the magnetic characterization curve shown in figure 7(d), note that in the figure of increment that the error in magnetic moment is around ± 9 $Am^2$/ kg in the region of 0.5 T. This may be happening because of the increase of noise ratio due to increased detection region or uncertainty in the modeling process.

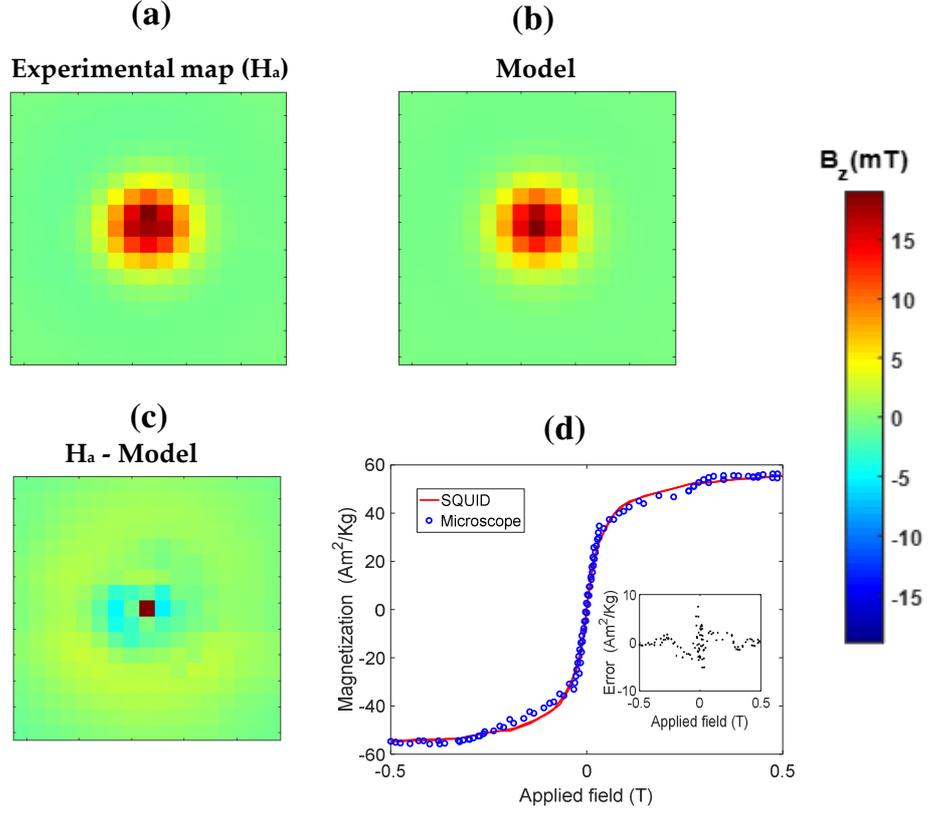

**Figure 7.** (a) Experimental mapping applying 20 mT. (b) Map obtained through the theoretical model of a current cylinder. (c) Subtraction of the experimental map by the theoretical. (d) Comparison of the magnetization results obtained using our technique (blue circles) and the values obtained for the same nanoparticle sample using a commercial magnetometer (MPMS SQUID, Quantum Design Inc.) (solid red curve).

5.3 Average diameter

In addition to characterizing the material behavior at room temperature, we estimated the average diameter of the NPs from the magnetization curves (see Fig. 6(d)) for the samples that exhibited superparamagnetic behavior at room temperature, as shown in Fig. 6(e). In general, the average diameter of NPs is obtained using techniques such as TEM and DLS. The average diameter of the NPs can be estimated from the magnetization curve measurements. In Eq. 3, we have [17,27]

$$D_{mag} = \frac{(18 k_B T \chi_0)^{1/3}}{(\mu_0 \pi M_S^2)^{1/3}} \qquad (3)$$

where $D_{mag}$ is the average diameter to be calculated, $T$ is room temperature, and $\chi_o$ is the initial susceptibility. Since the two samples are primarily composed of magnetite (see Fig. 3), we used a value of $\rho = 5.20 \times 10^3$ kg/m$^3$ [28]. Using the curves of the figure 6(d) for fields near zero, we can estimate the value of $\chi_o$. Hence, $M_s$ was estimated by extrapolating the magnetization curve as a function of the inverse field (1/H) for 1/H = 0 [15,17]. In this way, we obtained $M_s$ = 88 and 91 Am$^2$/kg for NPs produced by co-precipitation and PLA in water, respectively. Therefore, the diameters for the NPs obtained from the experimental magnetic data are 9 and 4 nm, respectively. These values are close to those calculated from the TEM images (11.6 nm and 4.4 nm, respectively; see Figs. 8(b) and (d)).

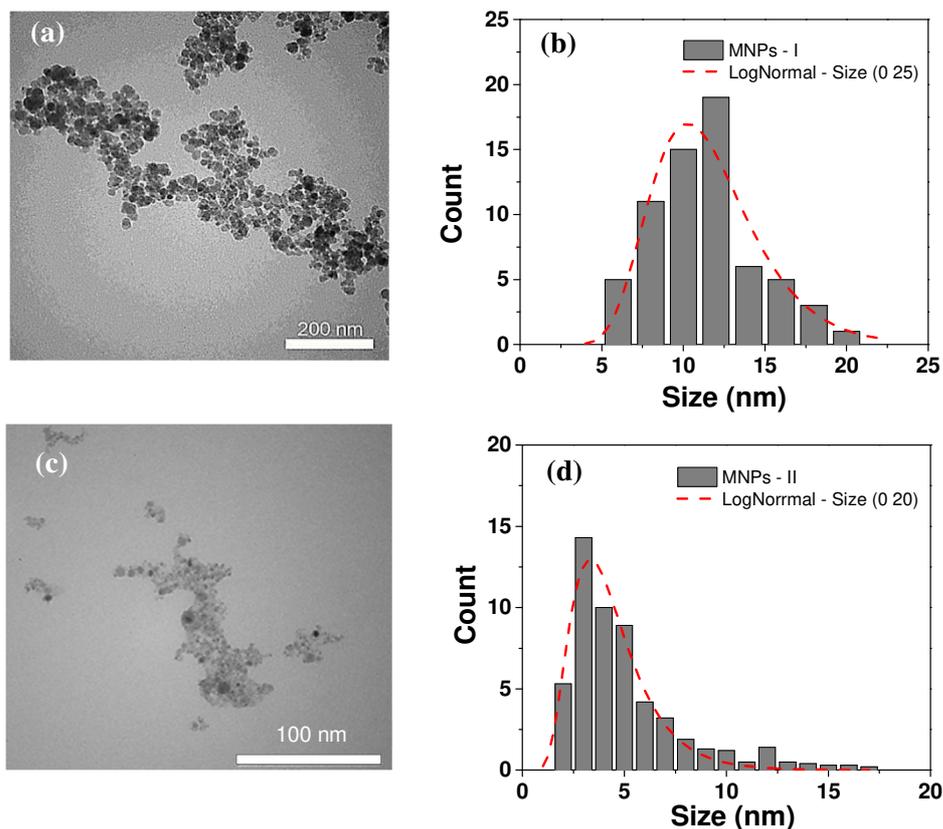

**Figure 8**. (a) TEM image of NPs produced by co-precipitation. (b) Histogram obtained from TEM images of NPs produced by co-precipitation. (c) TEM image of NPs produced by PLA. (d) Histogram obtained from TEM images of NPs produced by PLA.

A sample of approximately 200 NPs for each image was used to build the statistical size distributions, and the corresponding mathematical fit was performed using a log-normal distribution following the same procedure as reported in [29]. The results indicate that mean sizes of the NPs synthesized by the two different techniques are quite different. The mean sizes and standard deviation of the radius distributions are $<r>_{cp}$ = 11.6 nm and $<\delta>_{cp}$ = 3.4 nm, respectively, for the chemical synthesized NPs and $<r>_{PLA}$ = 4.4 nm, $<\delta>_{PLA}$ =1.9 nm, respectively, for the nanomaterial produced by PLA in water. The same trend is observable in the DLS measurements, as shown in Fig. 9.

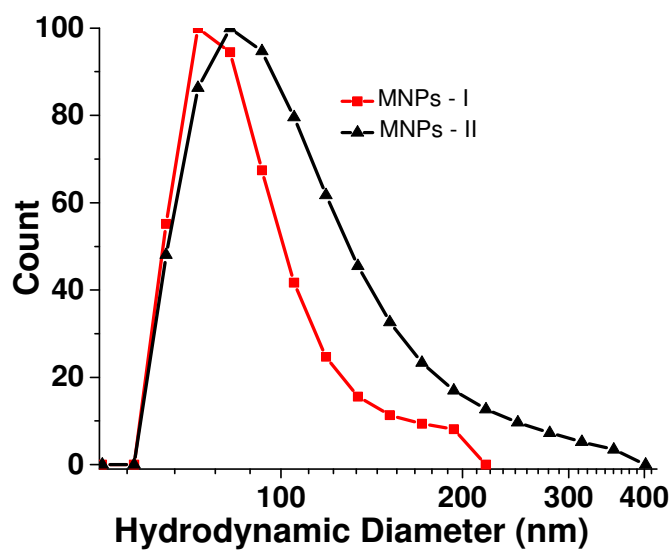

**Figure 9**. Size distributions of magnetic NPs synthesized by chemical routes (red line) and PLA in water (black line) obtained by the DLS technique.

The mean radius of the NPs synthesized chemically and by PLA are approximately 53 nm and 43 nm, respectively. As expected, the NPs sizes obtained by DLS are greater than those obtained by TEM, since the DLS technique measures the hydrodynamic radius of the NPs in a suspension, which is always larger than the metal nucleus size [20]. Table 1 presents the magnetic NPs (MNPs) sizes determined using different techniques. We attribute the differences to the intrinsic characteristics of each analysis method.

Table 1. Comparison of particle sizes obtained by various techniques.

| Technique | TEM | Magnetic | DLS |
|---|---|---|---|
| Co-precipitation | 12 nm | 9 nm | 106 nm |
| Laser ablation | 4 nm | 4 nm | 86 nm |

Therefore, we have demonstrated that the average nanoparticle size can be estimated, which is of practical interest. In addition, this estimation method is less expensive than the TEM technique.

## 6. Conclusion

The results obtained by magnetic measurements of NPs by SMM characterization using the proposed technique for obtaining magnetization curves were consistent and statistically comparable with those obtained by other instruments such as commercial magnetometers (SQUID). The errors obtained for the saturation and remanent magnetization were approximately ± 0.18 Am$^2$/kg and ± 0.6 Am$^2$/kg, respectively. Our findings demonstrate that this method is an excellent alternative for the magnetic characterization of NPs in micrograms samples. In addition, magnetic NPs produced by the different synthesis methods were shown to be uniform in terms of composition, in accordance with the results of DLS and TEM. These analyses also demonstrate that the characterization based on magnetic curves obtained by this new technique provides a reliable estimation of average particle size.

### Acknowledgments

This work was supported in part by the Brazilian agencies CNPq, Capes, Faperj and Fapesp. We thank Fredy Osorio and Amanda Farias dos Santos for Figure 4. (d). A special acknowledgment goes to and Dr. Wagner Wlysses for reviewing the manuscript.